\documentclass[twocolumn,prd,aps,showpacs,showkeys,amsmath,amssymb,nofootinbib]{revtex4-1}
\usepackage{bm}

\usepackage{color}
\usepackage{amsmath}
\usepackage{amsfonts}
\usepackage{verbatim}
\usepackage{amssymb}
\usepackage{graphicx}
\usepackage{epstopdf}
\usepackage{mathrsfs}
\usepackage{epsfig}
\usepackage{slashed}
\usepackage{bbold}
\usepackage{color} 

\newcommand{\bl}{\boldsymbol}

\newcommand{\ph}{\phantom}
\newcommand{\pd}{\partial}

\begin{document}


\title{On Conserved Quantities for the Free Motion of Particles with Spin}
\author{Esdras Barbosa dos Santos and Carlos Batista}
\email[]{carlosbatistas@df.ufpe.br, esdras.bsantos@df.ufpe.br}
\affiliation{Departamento de F\'{\i}sica, Universidade Federal de Pernambuco,
Recife, Pernambuco  50740-560, Brazil}


\begin{abstract}
In the early 80's, R. R\"{u}diger published a pair of articles in which it was found the most general conserved charges associated to the motion of particles with spin moving in curved spacetime. In particular, it was shown that besides the well-known conserved quantity associated to Killing vectors, it is also possible to have another conserved quantity that is linear in the spin of the particle if the spacetime admits a Killing-Yano tensor. However, in these papers it was proved that in order for this new scalar to be conserved two obscure conditions involving the Killing-Yano tensor and the curvature must be obeyed. In the present paper we try to shed light over these conditions and end up proving that this conserved quantity is useless for most physically relevant spacetimes. Notably, for particles moving in vacuum (Einstein spacetimes) this conserved scalar constructed with the Killing-Yano tensor will not help on the integration of the equations of motion. Moreover, we prove that, as a consequence of these obscure conditions, the Killing-Yano tensor must be covariantly constant.
\end{abstract}

\maketitle



\section{Introduction}

It is well-known that in Einstein's gravitational theory point particles that are free, i.e. interacting just with the gravitational field through the curved spacetime, move along geodesics. However, if the particle has a finite size, as all classical particles certainly do, it can carry internal angular momentum which, in turn, couple to the gravitational field and deviate the particle from the geodesic path. Indeed, energy can be stored in the form of angular momentum and, due to the Equivalence principle, any form of energy will respond to the gravitational field. Here we shall refer to this internal angular momentum that stems from the rotation of the particle around its own center of mass as ``spin''. The equations that dictate the motion of a test particle with spin are called Mathisson-Papapetrou-Dixon (MPD) equations and are given by \cite{Mathisson,Papapetrou,Dixon}
\begin{equation}\label{MPDeq}
  \left\{
    \begin{array}{ll}
      \dot{P}^\mu = -\frac{1}{2} R^\mu_{\ph{\mu} \nu\alpha\beta} V^\nu\,S^{\alpha\beta} \\
      \quad \\
      \dot{S}^{\alpha\beta} = P^\alpha\,V^\beta - V^\alpha\,P^\beta
    \end{array}
  \right.
\end{equation}
In these equations, $\bl{P}$ is the linear momentum of the particle, $S^{\alpha\beta}=S^{[\alpha\beta]}$ is its intrinsic angular momentum, whereas $V^\mu$ is the normalized velocity, $V^\mu V_{\mu} = 1$. The dot represents a covariant derivative along the movement of the particle. For instance, $\dot{P}^\mu = V^\nu \nabla_\nu P^\mu$. In particular, if a particle is point-like its moment of inertia vanishes, which in the classical realm implies that it cannot store energy in its spin. In the latter case we would have $\bl{S}=0$, so that the second equation above implies that $\bl{P}$ and $\bl{V}$ are proportional to each other, whereas the first equation yields that the movement is geodesic, as it should be. It is worth pointing out that these equations assume the so-called pole-dipole approximation, where multipoles of the energy-momentum tensor with order higher than one in the deviation from the center of mass are neglected. In the case of highly inhomogeneous gravitational fields and high spins, the quadrupole degree of freedom can be as important as the dipole term \cite{Semerak}.  For a nice sum up of some of the advances and lines of research on this topic up to the year 2007, see Ref. \cite{Kyrian:2007zz}. For a more recent review on the theme, the reader is referred to \cite{Book}.

If a spacetime admits a Killing vector field $\bl{K}$ then it follows that the scalar $P^\mu K_\mu$ is conserved along a geodesic vector field $\bl{P}$ with affine parametrization.  Likewise, if $N_{\mu\nu} = N_{(\mu\nu)}$ is a Killing tensor then the quadratic scalar in the momentum $P^\mu P^\nu N_{\mu\nu}$ is also conserved along geodesics. Moreover, if $Y_{\mu\nu}=-Y_{\nu\mu}$ is a Killing-Yano tensor then $N_{\mu\sigma} = Y_{\mu\nu}Y^{\nu}_{\ph{\nu}\sigma}$ is a Killing tensor, so that Killing-Yano (KY) tensors can be used to generate conserved charges along the geodesics that are quadratic on the momentum. These conserved charges are of central importance in the integration of the geodesic equation and, therefore, for obtaining the path followed by point-like free test particles. For instance, the geodesic motion on Kerr background can be fully integrated thanks to the existence of two Killing vector fields and a KY tensor \cite{Carter-constant,Walk-Pen}, see also \cite{Kubiz,Krtous} for higher-dimensional examples. However, it turns out that Killing tensors and KY tensors are also associated to the separability of field equations other than the geodesic equation, as Klein-Gordon, Maxwell, and Dirac equations \cite{Frol-KG, Oota}, as well as gravitational perturbation equation \cite{Teukolsky,OotaPerturb}. Therefore, it is natural to wonder whether Killing vectors and KY tensors lead to conserved charges for the movement dictated by Eq. (\ref{MPDeq}).

Concerning Killing vectors,  it is well-known that they yield conserved charges for the MPD equations. More precisely, if $\bl{K}$ is a Killing vector field then the scalar
$$ Q_{\bl{K}} = P^\mu\,K_\mu + \frac{1}{2}\,S^{\mu\nu}\nabla_\mu K_\nu  $$
is such that $\dot{Q}_{\bl{K}}=0$ whenever Eq. (\ref{MPDeq}) is assumed to hold. Regarding the role of Killing Tensors and KY tensors as generators of conserved charges for the MPD equations, some important results are not widespread on the literature. Important conclusions on this matter have been attained by R. R\"{u}diger in two articles published in early 80's \cite{Rudiger1,Rudiger2}. In spite of obtaining solid and useful results, these works got little attention and have very few citations. In particular, it has been proved there that if $\bl{Y}$ is a KY tensor then the scalar
\begin{equation}\label{CS}
  Q_{\bl{Y}} = S^{\alpha\beta}\widetilde{Y}_{\alpha\beta}
\end{equation}
can be a conserved charge for the MPD equations, where $\bl{\widetilde{Y}}$ stands for the Hodge dual of $\bl{Y}$, which is a closed conformal KY tensor. The interesting thing about the conserved scalar $Q_{\bl{Y}}$ is that it is linear in the angular momentum, whereas for the geodesic motion KY tensors are associated to conserved charges that are quadratic on the linear momentum. However, the conservation of $Q_{\bl{Y}}$ holds only if a pair of additional conditions involving $\bl{\widetilde{Y}}$, its derivative and the curvature are obeyed. However, as presented in Ref. \cite{Rudiger1}, these additional conditions are very obscure and their consequences have not been worked out so far. The aim of the present article is to shed light over those conditions. More precisely, by manipulating these additional constraints along with the integrability conditions necessary for the existence of KY tensors we will conclude that in most physical scenarios the charge $Q_{\bl{Y}}$ will be of no practical relevance.

The outline of the present article is the following. In Sec. \ref{Sec.ConservedCharge} it is shown the procedure used by R\"{u}diger to obtain the scalar $Q_{\bl{Y}}$ and the additional conditions necessary for it to be conserved. Some improvements on the deduction are done as well as a sign correction in one of R\"{u}diger's equation.  Then, in Sec. \ref{Sec.KY} KY tensors and their integrability conditions are reviewed. Sec. \ref{Sec.Main} then presents the main results of this article. There the additional conditions for the conservation of  $Q_{\bl{Y}}$ are worked out along with the integrability conditions necessary for the existence of a KY tensor. It is then obtained that the range of spacetimes such that $Q_{\bl{Y}}$ lead to a useful conserved charge is very narrow. Some examples of spacetimes allowing a nontrivial conserved scalar $Q_{\bl{Y}}$ are then found in Sec. \ref{Sec.Examples}. Finally, conclusions are summed up in Sec. \ref{Sec.Conclusions}.

Before proceeding, let us establish some notational conventions. Indices enclosed by round brackets are assumed to be symmetrized, whereas square brackets denote antisymmetrization of indices, so that $T_{(ab)} = (T_{ab}+T_{ba})/2$ and $T_{[ab]} = (T_{ab}-T_{ba})/2$; in addition, the tilde over a skew-symmetric tensor stands for the Hodge dual operation, $\widetilde{S}_{\mu\nu} = \frac{1}{2!}S^{\alpha\beta}\epsilon_{\alpha\beta\mu\nu}$. In what follows it is always assumed a four-dimensional spacetime endowed with a metric and the Levi-Civita connection.


\section{R\"{u}diger's Conserved Charge}\label{Sec.ConservedCharge}

In this section, we follow the steps adopted by R\"{u}diger in Ref. \cite{Rudiger1} in order to obtain the most general conserved scalar for the spinning particle that is linear in its momenta (linear and angular). However, in order to attain the desired result it is first necessary to digress about the supplementary condition required in order to complement MPD.

Note that the unknowns of Eq. (\ref{MPDeq}) comprise thirteen  degrees of freedom, four from $P^\mu$, six from $S^{\mu\nu}$ and three from $V^\mu$, since the velocity is assumed to be normalized, $V^{\alpha}V_{\alpha}=1$. However, MPD equations amount to ten constraints. Thus, three further constraints are necessary. The two most popular supplementary conditions adopted in the literature are the Pirani condition, defined by $S^{\alpha\beta}V_{\beta}=0$, and the Tulczyjew condition, defined by $S^{\alpha\beta}P_{\beta}=0$. Note that, although both conditions seem to impose four constraints, since there is one free index in these equations, the skew-symmetry of $S^{\alpha\beta}$ implies that just three directions of this free index yield actual constraints. For instance, projecting the constraint $S^{\alpha\beta}V_{\beta}=0$ in the direction $V_\alpha$ yields $0=0$, which represents no constraint. From the physical point of view, the non-uniqueness of the supplementary condition stems from the fact that our particle is assumed to be finite, so that there are an infinitude of points inside the body that one can use to define the orbit of the body, each choice lead to a different trajectory and a different velocity. However,  the orbits predicted by the several options are all close to each other for small spin, actually they are all contained in the world-tube of the particle \cite{Kyrian:2007zz}. The two choices mentioned above are just two popular ones due to the fact that the vectors $V^\alpha$ and $P^\alpha$ are naturally defined in the theory. Thus, in a sense, these choices do not break covariance. Following R\"{u}diger's choice, here we adopt Tulczyjew condition, namely $S^{\alpha\beta}P_{\beta}=0$. This is the usual choice for massive particles, since in flat spacetime it is associated to a unique trajectory, whereas Pirani's condition allows some freedom.

Assuming $S^{\alpha\beta}P_{\beta}=0$, it follows that $\mu^{2}=P^{\alpha}P_{\alpha}$ and $S^{\alpha\beta}S_{\alpha\beta}$ are both conserved along the orbit. But, for the goal of the present work, the most relevant feature of the latter condition is that it implies a relation in which the velocity is explicitly written in terms of the linear momentum $\bl{P}$ and spin vector $\bl{\Sigma}$, as we shall prove in Eq. \eqref{VPS}.

It is worth mentioning that for massless particles, however, the condition $S^{\alpha\beta}V_{\beta}=0$ seems to be more adequate, as argued, for example, in Refs. \cite{mashhoon,bailyn}. Beyond the two supplementary conditions above, a relaxed version of the Tulczyjew condition, $S^{\alpha\beta}P_{\beta} \propto P^{\alpha}$, was proposed in \cite{armaza} for massless particles.

Contracting the spin equation in (\ref{MPDeq}) with $P_{\beta}$ and using the derivative of the supplementary condition $S^{\alpha\beta}P_{\beta}=0$, we eventually arrive at
\begin{equation}\label{7}
V^{\nu}=\frac{m}{\mu^{2}}P^{\nu}-\frac{1}{2 \mu^{2}}S^{\nu\alpha}R_{\alpha\beta\gamma\delta}V^{\beta}S^{\gamma\delta},
\end{equation}
where $m=P_\mu V^\mu$ and $\mu^2 = P^\alpha P_\alpha$. Then, defining 
$$ a\equiv\frac{m}{\mu^{2}}\;,\;\; \text{ and }\; D^\nu_{\ph{\nu}\beta} \equiv -\frac{1}{2 \mu^{2}}\,S^{\nu\alpha}R_{\alpha\beta\gamma\delta}S^{\gamma\delta} \,,$$
it follows that the above expression can be written as
$$ V^{\nu}=a\,P^{\nu} +  D^\nu_{\ph{\nu}\beta} \,V^\beta \,.$$
This expression can be iterated by inserting itself in the right hand side, so that we end up with
\begin{align}
  V^\nu &= a\,P^\nu + D^\nu_{\ph{\nu}\beta}\left[ aP^\beta + D^\beta_{\ph{\beta}\sigma} V^\sigma  \right] \nonumber\\
  &= a\,P^\nu + D^\nu_{\ph{\nu}\beta}\left[ aP^\beta + D^\beta_{\ph{\beta} \sigma}\left(a P^\sigma + D^\sigma_{\ph{\sigma}\rho} V^\rho  \right) \right] \nonumber\\
&= a\,\left[ P^\nu + D^\nu_{\ph{\nu}\beta_1} P^{\beta_1} + D^\nu_{\ph{\nu}\beta_1} D^{\beta_1}_{\ph{\beta_1}\beta_2} P^{\beta_2} + \cdots \right]
\label{Series}
\end{align}
Thus, we have just found an expression for $\bl{V}$ in terms of the momenta, although in the form of an infinite series. It would be nice to sum this series and attain a finite formula on the right hand side, as we shall do in the sequel.

The supplementary condition $S^{\alpha\beta}P_\beta=0$ implies that there exists some vector $\Sigma^\mu$, dubbed spin 4-vector, such that 
\begin{equation}\label{SSigma}
  S^{\alpha\beta}=\epsilon^{\alpha\beta\mu\nu} \Sigma_{\mu}P_{\nu}\,.
\end{equation}
Moreover, since the transformation $\Sigma^\mu \rightarrow \Sigma^\mu + \lambda P^\mu$ does not change the above expression for $S^{\alpha\beta}$, for an arbitrary $\lambda$, and since we are assuming that $\mu^2=P^\mu P_\mu \neq 0$, it follows that we can impose that $\Sigma^\mu P_\mu = 0$. This imposition represents not loss of generality. Thus, $\bl{\Sigma}$ has three degrees of freedom. One can then prove that  $S^{[\alpha\beta}S^{\mu]\nu}=0$ holds as a consequence of Eq. (\ref{SSigma}). Then, using the latter relation, one can establish that
$$ D^\nu_{\ph{\nu}\beta_1} D^{\beta_1}_{\ph{\beta_1}\beta_2} P^{\beta_2} =\frac{1}{2}\, D^\nu_{\ph{\nu}\beta_1} P^{\beta_1} D^{\beta_2}_{\ph{\beta_1}\beta_2} = \frac{d}{2}\, D^\nu_{\ph{\nu}\beta_1} P^{\beta_1}\,, $$
where $d\equiv D^{\beta}_{\ph{\beta}\beta}$. From the latter relation, it then follows that
$$ D^\nu_{\ph{\nu}\beta_1} D^{\beta_1}_{\ph{\beta_1}\beta_2} D^{\beta_2}_{\ph{\beta_1}\beta_3} P^{\beta_3} = \left(\frac{d}{2} \right)^2\, D^\nu_{\ph{\nu}\beta_1} P^{\beta_1} \,. $$
Thus, the series (\ref{Series}) can be written as
\begin{align}\label{VPS}
  V^\nu &= a\,\left[ P^\nu + D^\nu_{\ph{\nu}\beta} P^{\beta}\,\sum_{n=0}^\infty \frac{ d^n}{2^n} \right]  \nonumber\\
  &= a\,P^\nu + \frac{a\, D^\nu_{\ph{\nu}\beta} P^{\beta}}{1-d/2}\,. 
\end{align}
Inserting the definitions of  $\bl{D}$ and $d$, we finally arrive at the desired relation. Thus, for the supplementary condition $S^{\alpha\beta}P_{\beta}=0$, it follows that the basic degrees of freedom are $P^\alpha$, $\Sigma^\alpha$, and $m\equiv P^\alpha V_{\alpha}$. In Ref. \cite{Rudiger1} the same relation has been attained by using a particular reference frame, whereas here no covariance breaking was necessary. Analogously, an expression for the velocity in terms of the momentum and the spin tensor can also be attained for the Pirani's supplementary condition, as recently proved in Ref. \cite{Costa}. The latter relation turns out to be equivalent to Pirani's supplementary condition, so that when we substitute the spin tensor in terms of the spin vector we end up with a trivial identity. This is the reason why such a relation would hardly be helpful for finding the
conserved quantities of MPD equation as we do in the sequel. However, the relation found in Ref. \cite{Costa} proved to be valuable on the integartion of MPD equations.

Once established Eq. (\ref{VPS}), we are ready to look  for the conserved charges following the steps of Ref. \cite{Rudiger1}. The most general scalar that is linear in the momenta is given by
\begin{equation}
Q=K_{\mu}P^{\mu}+L_{\mu\nu}S^{\mu\nu},\label{5}
\end{equation}
for some tensors $K_{\mu}$ and $L_{\mu\nu} = L_{[\mu\nu]}$. Now, let us impose  that $Q$ is conserved along particle's trajectory and then verify what conditions this requirement implies for the tensors $\bl{K}$ and $\bl{L}$. More explicitly, assuming $\dot{Q}=0$ it follows that  
\begin{multline}
V^{\alpha}(\nabla_{\alpha}K_{\mu}P^{\mu}
-\frac{1}{2}K_{\mu}R^{\mu}\hspace{0.025cm}_{\alpha\beta\gamma}S^{\beta\gamma}\\
+\nabla_{\alpha}L_{\mu\nu}S^{\mu\nu}+2 L_{\mu\alpha}P^{\mu})=0\,,\label{Qdot}
\end{multline}
where MPD equations have been used. The next step is replacing $V^\alpha$ in terms of the momenta, by means of (\ref{VPS}) and finally write $S^{\alpha\beta}$ in terms of the spin vector $\Sigma^\alpha$, so that the supplementary condition is already taken into account. Doing so, we end up with a relation containing just the fundamental degrees of freedom, namely $P^\alpha$, $\Sigma^\alpha$, and $m$. The next step is to impose that Eq. (\ref{Qdot}) holds for arbitrary values of these independent degrees of freedom. This was the procedure adopted by R\"{u}diger in Refs. \cite{Rudiger1,Rudiger2}. The dependence on $m$ is not relevant, since it factors out as a collective multiplicative factor. Since $P^\alpha$ and $\Sigma^\alpha$ are independent of each other and arbitrary, terms with different powers of these degrees of freedom must vanish independently. For instance, the unique term in Eq. (\ref{Qdot}) that is of order two in $\bl{P}$ and of order zero in $\bl{\Sigma}$ is $\nabla_{\alpha}K_{\beta}P^{\alpha}P^{\beta}$, so that we can conclude that
$$ \nabla_{\alpha}K_{\beta}P^{\alpha}P^{\beta} = 0 $$
for an arbitrary $\bl{P}$. The latter condition, in turn, implies that $\nabla_{(\alpha}K_{\beta)}=0$, i.e. $\bl{K}$ is a Killing vector field. Likewise, the unique term of order two in $\bl{P}$ and order one in $\bl{\Sigma}$ yields
\begin{equation}
\epsilon^{\mu\nu\gamma}_{\ph{\mu}\ph{\nu}\ph{\gamma} (\alpha}\nabla_{\beta)}\left(L_{\mu\nu}-\frac{1}{2}\nabla_{\mu}K_{\nu}\right)=0 \,,\label{10}
\end{equation}
where the identity $\nabla_{\alpha}\nabla_{\beta}K_{\gamma}=K_{\mu}R^{\mu}_{\ph{\mu} \alpha\beta\gamma}$ has been used, which stems from the fact that $\bl{K}$ is a Killing vector field. Thus, defining
$$ Y_{\alpha\beta} \equiv \frac{1}{2}\,\epsilon^{\mu\nu}_{\ph{\mu}\ph{\nu}\alpha\beta}\left(L_{\mu\nu}-\frac{1}{2}\nabla_{\mu}K_{\nu}\right) \,, $$
it follows from Eq. (\ref{10}) that $\bl{Y}$ must obey the equation $\nabla_{(\alpha}Y_{\beta)\nu}=0$, i.e. it must be a Killing-Yano tensor. Thus, the tensor $L_{\mu\nu}$ must be written as $L_{\mu\nu} = \frac{1}{2}\nabla_{\mu}K_{\nu} + \widetilde{Y}_{\mu\nu}$. Two other conditions can be extracted from Eq. (\ref{Qdot}), one that comes from a term of order four in $\bl{P}$ and order two in $\bl{\Sigma}$, while the other is of order four in $\bl{P}$ and order three in $\bl{\Sigma}$. These two conditions are respectively given by
\begin{align}
  \left[\tilde{\widetilde{R}}^{\kappa(\alpha\beta)}_{\ph{k}\ph{a}\ph{b}\;(\mu} g_{\nu\rho}+\tilde{\widetilde{R}}^{\kappa\ph{\mu}\ph{\nu}(\alpha}_{\ph{k}(\mu\nu}\delta^{\beta)}_{\rho}\right]
\widetilde{Y}_{\sigma) \kappa}=0, \label{11}\\
  \left[\tilde{\widetilde{R}}^{\kappa(\alpha\beta}_{\ph{k}\ph{a}\ph{b}\; (\mu}g_{\nu\rho|}+\tilde{\widetilde{R}}^{k\ph{d}\ph{e} (\alpha}_{\ph{k}(\mu\nu}\delta^{\beta}_{\rho|}\right]  \nabla_{\kappa}Y^{\gamma)}_{\ph{\gamma}\vert \sigma)}=0. \label{12}
\end{align}
where we have introduced the double Hodge dual of the Riemann tensor. More precisely, here we shall adopt the following definitions: 
\begin{align*}
  \tilde{\widetilde{R}}_{\alpha\beta\mu\nu}&=\frac{1}{4}\epsilon_{\alpha\beta\alpha'\beta'} R^{\alpha'\beta'\mu'\nu'} \epsilon_{\mu'\nu'\mu\nu}\,, \\
  \widetilde{R}_{\alpha\beta\mu\nu}&=\frac{1}{2}\epsilon_{\alpha\beta\alpha'\beta'} R^{\alpha'\beta'}_{\ph{\alpha'\beta'}\mu\nu}  \,.
\end{align*}
As pointed out by R\"{u}diger in Ref. \cite{Rudiger1}, Eqs. (\ref{11}) and (\ref{12}) can be simplified. Indeed, after some algebra, one can prove that they are equivalent to the following two constraints respectively:
\begin{align} 
  \tilde{\widetilde{R}}^{\sigma(\alpha\beta)}_{\ph{k}\ph{a}\ph{b}\, (\gamma}\widetilde{Y}_{\delta)\sigma}-\frac{1}{6}G^{\sigma(\alpha}\delta^{\beta)}_{(\gamma}\widetilde{Y}_{\delta)\sigma}+
\frac{1}{4}\widetilde{Y}_{\rho\sigma}\tilde{\widetilde{R}}^{\rho\sigma(\alpha}_{\ph{h}\ph{k}\ph{a}(\gamma}
\delta^{\beta)}_{\delta)}=0 \,, \label{13}\\
\quad\nonumber \\
  J^{(\alpha}\widetilde{R}^{\beta\ph{f}\ph{e}\; \gamma)}_{\ph{b}(\mu\nu)}-J^{\kappa}\widetilde{R}_{\kappa(\mu\nu)}^{\ph{k}\ph{f}\ph{e}\; (\alpha}g^{\beta\gamma)}-J^{\kappa}\widetilde{R}_{\kappa\ph{a}\ph{b}(\mu}^{\ph{k}(\alpha\beta}\delta^{\gamma)}_{\nu)}=0 \,, \label{14}
\end{align}
where $J^{\alpha}$ is the divergence of $\bl{\widetilde{Y}}$, namely $J_\beta = \nabla^\alpha \widetilde{Y}_{\alpha\beta}$, whereas $G_{\mu\nu}$ stands for the Einstein tensor. We note, however, that there is a sign difference between our Eq. (\ref{13}) and Eq. (4.9) of Ref. \cite{Rudiger1}, in the latter the sign in front of the fraction $1/6$ is positive, although the correct sign is negative, as written here. Indeed, should the sign be positive such constraint would not be identically valid for maximally symmetric spacetimes, as it should be, as acknowledged by R\"{u}diger himself. Thus, there must have been a typo at this point in Ref. \cite{Rudiger1}.

Summing up, assuming Tulczyjew supplementary condition, we have proved that the most general conserved charged for MPD equations that is linear in momenta is given by
\begin{equation}
Q= \left( K_{\mu}P^{\mu}+\frac{1}{2}\nabla_{\mu}K_{\nu}S^{\mu\nu}\right) \,+\, \widetilde{Y}_{\mu\nu}S^{\mu\nu},\label{16}
\end{equation}
where $\bl{K}$ is a Killing vector and $\bl{Y}$ is a rank two KY tensor. In addition the constraints (\ref{13}) and (\ref{14}) must hold. Since $\bl{K}$ and $\bl{Y}$ are totally independent from each other and the latter constraints do not depend on $\bl{K}$, it follows that the scalars 
$$ Q_{\bl{K}} =  K_{\mu}P^{\mu}+\frac{1}{2}\nabla_{\mu}K_{\nu}S^{\mu\nu} \;\; \text{and}\;\;  Q_{\bl{Y}} = \widetilde{Y}_{\mu\nu}S^{\mu\nu} \,, $$
are independently conserved. Indeed, it is widely known that $Q_{\bl{K}}$ is conserved for any Killing vector $\bl{K}$. The important result of Ref. \cite{Rudiger1} is that the scalar $Q_{\bl{Y}}$ is conserved as long as $\bl{Y}$ is a KY tensor and conditions  (\ref{13}) and (\ref{14}) hold. The problem is that the latter conditions are quite obscure and have not been tackled in the literature so far. The main goal of the present work is to shed light over the meaning of these constraints and determine the scenarios in which the conserved charge $Q_{\bl{Y}}$ is allowed to exist.

%

\section{Killing-Yano Tensors and its Integrability Conditions}\label{Sec.KY}

A Killing tensor is a totally symmetric tensor $N_{\alpha_1\cdots \alpha_p} = N_{(\alpha_1\cdots\alpha_p)}$ that obeys the equation $\nabla_{(\beta}N_{\alpha_1\cdots\alpha_p)}=0$. In particular, Killing vectors can be seen as Killing tensors of rank one. Just as Killing vector fields generate symmetries on the spacetime, which therefore lead to conservation laws for the geodesic motion, Killing tensors are the generators of symmetries on the phase space of the geodesic Hamiltonian and, due to N\"{o}ther's theorem, also yield conserved charges along geodesics \cite{Santillan}. Since Killing tensors are not related to symmetries of the spacetime itself, they are referred to as hidden symmetries and are generally more hard to find than Killing vector fields. Indeed, it took a while to perceive that Kerr solution is endowed with a Killing tensor in addition to the two Killing vector fields associated to stationarity and axial symmetry. The Killing tensor of Kerr spacetime was the missing link necessary to attain full integrability for the orbits of point-like test particles moving in this background \cite{Carter-constant,Walk-Pen}.

Another important mathematical object for these matters are the Killing-Yano (KY) tensors, which are totally skew-symmetric tensors, $Y_{\alpha_1\cdots\alpha_q}=Y_{[\alpha_1\cdots\alpha_q]}$, that obey equation $\nabla_{(\beta}Y_{\alpha_1)\alpha_2\cdots\alpha_q}=0$, which is also a generalization of the Killing vector equation. It turns out that the square of a KY tensor is always a Killing tensor of rank two, $N_{\mu\nu}= Y_\mu^{\ph{\mu}\alpha_2\cdots\alpha_q}Y_{\nu\alpha_2\cdots\alpha_q}$. Nevertheless, it is worth mentioning that rank two Killing tensors are not necessarily the square of a KY tensor, just in special cases this turn out to be true \cite{Collinson,Steph_KY}. Thus, one can say that KY tensors are more special than Killing tensors. Indeed, in addition to generating conserved charges along the geodesic motion, via the Killing tensor built from its square, KY tensors are also related to symmetries of the phase space of a semi-classical supersymmetric model for free particles with quantum spin $1/2$ whose internal angular momentum is represented by  $S^{\alpha\beta}=\xi^\alpha\xi^\beta$, where $\xi^\alpha$ is a Grassmann variable \cite{SpinningGrasmann,SpinningGrasmann2,Santillan}. Furthermore, KY tensors can be used to construct operators that commute with the D'Alembertian and the Dirac operators \cite{Benn-DiracSymme,Cariglia}, which is of relevance to describe quantum particles moving in classical spacetimes. It is said that KY symmetries are non-anomalous, a feature that generally is not shared by the Killing tensors. KY tensors have also been used to build Lax pairs in curved spaces \cite{KY-Lax}, which is of relevance for the theory of integrable systems.

Suppose that $Z^\mu$ is a covariantly constant vector field, namely $\nabla_\mu Z_\nu = 0$. Then, using this hypothesis along with Ricci identity it follows that 
$$ 0 = 2\nabla_{[\mu}\nabla_{\nu]}Z^\alpha = R^\alpha_{\ph{\alpha}\beta\mu\nu}Z^\beta \,. $$
The latter equation is said to be an integrability condition for the existence of a constant vector field. For instance, if the curvature of a connection is such that there exists no direction $T^\alpha$ obeying $T^{\alpha}R_{\alpha\beta\mu\nu}=0$, then we can already state that no covariantly constant vector field exists, without needing to bother about integrating the differential equation $\nabla_\mu Z_\nu = 0$ for a generic vector field $Z^\mu$. Likewise, in order to enable a KY tensor to exist in a spacetime some integrability conditions must hold. For instance, concerning KY tensors of rank two, $Y_{\mu\nu}$, the following constraints must hold \cite{carlosIntCond,Kashiwada-Int,Tachibana-KY}:
\begin{align}
  0 =& R^{\beta}_{\ph{\beta}(\mu}\,Y_{\nu)\beta} \,, \label{Integ.Cond1}  \\
  0=& C_{\alpha\beta[\mu}^{\ph{\alpha\beta[\mu}\sigma} \,Y_{\nu]\sigma} +C_{\mu\nu[\alpha}^{\ph{\alpha\beta[\mu}\sigma} \,Y_{\beta]\sigma} \,, \label{Integ.Cond2}
\end{align}
where $R^{\beta}_{\ph{\beta}\mu}$ stands for the Ricci tensor whereas $C_{\mu\nu\alpha\beta}$ denotes the Weyl tensor. Thus, the curvature of the spacetime must obey some algebraic restrictions if a spacetime admits a KY tensor. We shall return to this point later, after introducing the basics of Petrov classification.

At this point it is useful to use a null tetrad frame $\{\bl{\ell},\bl{n},\bl{m},\bar{\bl{m}}\}$, where $\bl{\ell}$ and $\bl{n}$ are real vector fields, whereas $\bl{m}$ is complex with $\bar{\bl{m}}$ being its complex conjugate. These reality conditions encode the fact that we are considering Lorentzian signature. By definition of a null tetrad frame, the only nonvanishing inner products in this frame are the following:
$$ \ell^\mu\,n_\mu = 1 \;\; \text{ and } \;\;  m^\mu\,\bar{m}_\mu = -1 \,.$$
In particular, all vectors of the frame are light-like. For instance, if $\{\bl{e}_0,\bl{e}_1, \bl{e}_2,\bl{e}_3\}$ is a Loretnz frame, with their inner products yielding the Minkowski metric then 
$$ \bl{\ell} = \frac{1}{\sqrt{2}}\left( \bl{e}_0 + \bl{e}_1\right) \,,\;\; \bl{n} = \frac{1}{\sqrt{2}}\left(  \bl{e}_0 - \bl{e}_1\right)\,,  $$  
$$ \bl{m} = \frac{1}{\sqrt{2}}\left( \bl{e}_2 + i\bl{e}_3\right) \,,\;\; \bar{\bl{m}} = \frac{1}{\sqrt{2}}\left(  \bl{e}_2 - i\bl{e}_3\right)\,,  $$
is a null tetrad frame. 
This kind of frame is valuable to define the components of the Weyl tensor in a compact way. The ten degrees of freedom of the Weyl tensor in four dimensions can be written in terms of five complex scalars known as Weyl scalars and defined by
\begin{equation}\label{weylscalars}
   \begin{array}{cc}
     \Psi_0 \equiv C_{\ell m \ell m} \; ,\; \; \Psi_1 \equiv C_{\ell n \ell m}\;, \; \; \Psi_2 \equiv C_{\ell m \bar{m} n}  \\
   \Psi_3 \equiv C_{\ell n \bar{m}n}\;,\;\; \Psi_4 \equiv C_{n \bar{m} n \bar{m}}\;,
   \end{array}
 \end{equation}
where in the above equation $C_{\ell n \ell m}$ is just a compact way of denoting $C_{\mu\nu\alpha\beta}\ell^\mu n^\nu \ell^\alpha m^\beta$ and so on.  The Petrov classification, an algebraic classification for the Weyl tensor that proved to be valuable in several physical and mathematical problems, can then be defined in terms of the vanishing of these Weyl scalars \cite{Bat-Book}. The table \ref{TablePetrov} summarizes such link. 
\begin{table}[ht!!]
\vspace{0.6cm}
\begin{center}
\begin{tabular}{c|c}
  \textbf{Petrov Type} & \textbf{Vanishing Weyl Scalars} \\ \hline\hline
  $I$ & $\Psi_0\,,\;\Psi_4$ \\ \hline
  $II$ & $\Psi_0\,,\;\Psi_1\,,\;\Psi_4$ \\ \hline
 $III$ &  $\Psi_0\,,\;\Psi_1,\;\Psi_2\,,\;\Psi_4$ \\ \hline
  $D$&  $\Psi_0\,,\;\Psi_1,\;\Psi_3\,,\;\Psi_4$ \\ \hline
  $N$ & $\Psi_0\,,\;\Psi_1,\;\Psi_2\,,\;\Psi_3$ \\ \hline
$O$ & $\Psi_0\,,\;\Psi_1,\;\Psi_2\,,\;\Psi_3 ,\;\Psi_4$ \\ \hline
  \hline
\end{tabular}
\end{center}
\caption{Petrov types and its relation with the possibility of annihilating the Weyl scalars by a judicious choice of null tetrad frame.  Note that the type $O$ means a conformally flat spacetime, i.e. the Weyl tensor is identically zero in such a case. }\label{TablePetrov}
\end{table}
For instance, if the Weyl tensor of a spacetime is of Petrov type $N$ then it is possible to find a null tetrad frame in which all Weyl scalars except $\Psi_4$ vanish. For a review on Petrov classification see \cite{Bat-Book} and references therein.

Null tetrad frames are also of relevance to define the possible algebraic types of a bivector, i.e. a rank two skew-symmetric tensor $B_{\mu\nu}=B_{[\mu\nu]}$.  In a four-dimensional Lorentzian space, any nonzero bivector can be of two algebraic types. Either it is a null bivector, meaning that both contractions $B^{\mu\nu}B_{\mu\nu}$ and $B^{\mu\nu}\widetilde{B}_{\mu\nu}$ vanish, or it is non-null. It turns out that given a real bivector $B_{\mu\nu}$ one can always find a null frame in which the bivector  is written in one of the following forms depending on its algebraic type:
\begin{equation}\label{BivectorTypes}
 \left\{
  \begin{array}{ll}
   \textrm{Null Bivector: }\bl{B} = \bl{\ell}\wedge (\bl{m} + \bar{\bl{m}}) \\
    \quad \\
    \textrm{Non-Null Bivector: }\bl{B} = f \,\bl{\ell}\wedge \bl{n} + i h  \,\bl{m} \wedge \bar{\bl{m}}\,,
  \end{array}
\right.
\end{equation}
where $f$ and $h$ are real functions that cannot vanish simultaneously. Since a rank two KY tensor is a bivector, we can then work out the consequences of the integrability condition (\ref{Integ.Cond1}) for the Ricci tensor. Actually, in the next section we will be more interested in the trace-less part of the Ricci tensor, which is defined by
$$ \Phi_{\mu\nu} = R_{\mu\nu} - \frac{1}{4}\,R\,g_{\mu\nu}\,, $$
where $R$ stands for the Ricci scalar, $R^\alpha_{\ph{\alpha}\alpha}$, and $g_{\mu\nu}$ is the metric. A spacetime is called an Einstein spacetime whenever its Ricci tensor is proportional to the metric, which is equivalent to say that $\bl{\Phi}$ vanishes. Note that, in terms of the null tetrad frame, the traceless condition implies that $\Phi_{\ell n} = \Phi_{m\bar{m}}$.

Now, assuming that the KY tensor is a null bivector, i.e. $\bl{Y} = \bl{\ell}\wedge (\bl{m} + \bar{\bl{m}})$ for some null frame, then inserting this form into Eq. (\ref{Integ.Cond1}), and finally contracting the free indices of this equation with the vectors of the null tetrad we eventually conclude that
\begin{equation}\label{RicciNull}
 \bl{Y} \textrm{ Null:} \left\{
     \begin{array}{ll}
       \Phi_{\ell\ell}=\Phi_{\ell m}=\Phi_{\ell \bar{m}}= \Phi_{nm} + \Phi_{n \bar{m}}=0\,, \\
       \Phi_{mm} =\Phi_{\bar{m} \bar{m}}=-2\Phi_{\ell n}\,.
     \end{array}
   \right.
\end{equation}
In the same fashion, assuming that the KY tensor is non-null and writing it in the standard form given in Eq. (\ref{BivectorTypes}), it follows that the integrability condition \eqref{Integ.Cond1} implies
\begin{equation}\label{RicciNonNull}
 \bl{Y} \textrm{ Non-Null:} \left\{
     \begin{array}{ll}
       \Phi_{\ell m}=\Phi_{\ell \bar{m}}= \Phi_{nm}= \Phi_{n \bar{m}} =0\,, \\
       \left\{
         \begin{array}{ll}
           f\neq 0 \;\Rightarrow\;\Phi_{\ell\ell} =\Phi_{nn}=0\,\\
           h\neq 0 \;\Rightarrow\;\Phi_{mm} =\Phi_{\bar{m}\bar{m}}=0\,.
         \end{array}
       \right.
     \end{array}
   \right.
\end{equation}
Thus, for a generic non-null KY, i.e. when the real functions $f$ and $h$ appearing in the standard form of Eq. (\ref{BivectorTypes}) are both nonvanishing, we have that $\Phi_{\ell\ell}$, $\Phi_{nn}$, $\Phi_{mm}$, and $\Phi_{\bar{m}\bar{m}}$ all vanish. However, if $h$ vanishes then we cannot assert that $\Phi_{mm} =\Phi_{\bar{m}\bar{m}}=0$, whereas if $f$ vanishes the integrability condition does not implies $\Phi_{\ell\ell} =\Phi_{nn}=0$. Recall that $f$ and $h$ cannot vanish simultaneously, otherwise the KY tensor would be trivial.

In the same vein, it is interesting to see the interplay between the Petrov classification and the possible algebraic types of a KY tensor. Assuming that $Y_{\mu\nu}$ is a KY whose algebraic type is null it follows that there exists some null frame such that $\bl{Y} = \bl{\ell}\wedge (\bl{m} + \bar{\bl{m}})$. Then, inserting this expression for the KY tensor into the integrability condition (\ref{Integ.Cond2}) it follows, after some algebra, that in this null frame the following Weyl scalars $\Psi_0$, $\Psi_1$, $\Psi_2$, and $\Psi_3$ must all vanish. Thus, for a null KY tensor the Petrov classification must be type $N$ or type $O$, where the latter is a degenerate case of type $N$ \cite{carlosIntCond}. Likewise, assuming that the KY tensor is non-null and inserting its generic form given in Eq. (\ref{BivectorTypes}) into the integrability condition (\ref{Integ.Cond2}), it follows that the unique Weyl scalar that can be different from zero is $\Psi_2$, so that the Petrov type is $D$ or $O$ (which is a degenerate case of $D$). Summing up, the following conclusion holds
\begin{equation}\label{KYWeylscalars}
 \left\{
  \begin{array}{ll}
   \textrm{Null KY: } \Psi_0=\Psi_1=\Psi_2=\Psi_3=0\,,  \\
    \quad \\
    \textrm{Non-Null KY: }\Psi_0=\Psi_1=\Psi_3=\Psi_4=0\,.
  \end{array}
\right.
\end{equation}
Thus, just from the algebraic type of the Weyl tensor one can already rule out the possible existence of a KY tensor of rank two. For instance, suppose that a spacetime is of Petrov type $III$, then it cannot admit a KY tensor. This statement can be done prior to any attempt of integrating the KY equation. Thus, the integrability conditions can be a very powerful tool. In the next section we shall use this tool along with the conditions (\ref{13}) and (\ref{14}), that are required in order to guarantee that the scalar $Q_{\bl{Y}}=S^{\alpha\beta}\widetilde{Y}_{\alpha\beta}$ is conserved along a solution of MPD equations, and conclude that very few spacetimes allow this conserved charge. In particular, we will prove that this scalar is useless for an Einstein spacetime.


\section{Spacetimes Allowing the Conserved Charge}\label{Sec.Main}

In this section we shall investigate the constraints (\ref{13}) and (\ref{14}) that are required to hold in order to guarantee that the scalar $Q_{\bl{Y}}$ is conserved. The idea is to study its consequences along with the integrability conditions that must be true due to the fact that $\bl{Y}$ is a KY tensor. As we will prove in the sequel, when analysed together, these constraints are very restrictive, with a very narrow class of spacetimes obeying them. Before proceeding, however, let us establish that for the maximally symmetric spacetimes, i.e. de Sitter, anti-de Sitter and Minkowski spacetimes, the conserved charge $Q_{\bl{Y}}$ is useless. This is a consequence of the fact that in these spaces the number of independent Killing vector fields is ten, leading to ten conserved charges $Q_{\bl{K}}$, which are enough to obtain expressions for the ten unknowns $\bl{P}$ and $\bl{S}$ in terms of the initial conditions of the particle. Since here we are assuming the supplementary condition $S^{\alpha\beta}P_{\beta}=0$, one can then use Eq. (\ref{VPS}) in order to obtain an expression for the velocity $\bl{V}$.  In fact, the full integrability of MPD equations for de Sitter spacetime has been explicitly attained in Ref. \cite{Obukhov:2010kn}. Thus, in this sense, one can say that the conservation of $Q_{\bl{Y}}$ is somehow trivial for maximally symmetric spacetimes, reason why we shall ignore this case in what follows. 


The Riemann tensor can be decomposed in terms of its irreducible blocks with respect to the action of the Lorentz group, which are the Weyl tensor, the trace-less part of the Ricci tensor and the Ricci scalar. This decomposition is explicitly written as 
\begin{equation}
R_{\alpha\beta\gamma\delta}=C_{\alpha\beta\gamma\delta}+g_{\alpha[\gamma}\Phi_{\delta] \beta}-g_{\beta[\gamma}\Phi_{\delta] \alpha}+\frac{R}{6}g_{\alpha[\gamma}g_{\delta]\beta}\,.
\end{equation}
In particular, the spacetime is maximally symmetric if, and only if, $C_{\mu\nu\alpha\beta}$ and $\Phi_{\mu\nu}$ vanish simultaneously. 
Each of the irreducible blocks have a simple transformation with respect to the double Hodge dual. More precisely, we have 
\begin{equation*}
\tilde{\widetilde{R}}_{\alpha\beta\gamma\delta}=-C_{\alpha\beta\gamma\delta}+g_{\alpha[\gamma}\Phi_{\delta] \beta}-g_{\beta[\gamma}\Phi_{\delta] \alpha}-\frac{R}{6}g_{\alpha[\gamma}g_{\delta]\beta} \,. \label{17}
\end{equation*}
Using this expression along with $G_{\alpha\beta}=\Phi_{\alpha\beta}-\frac{R}{4} g_{\alpha\beta}$, it follows that the constraint (\ref{13}) can be equivalently written as
\begin{align} 
  &C^{\kappa(\alpha\beta)}_{\ph{\kappa}\ph{\alpha}\ph{\beta}\; (\gamma}\widetilde{Y}_{\delta)\kappa}+ \frac{1}{4} \widetilde{Y}_{\epsilon\kappa} C^{\epsilon\kappa(\alpha}_{\ph{h}\ph{\kappa}\ph{\alpha}(\gamma}\delta^{\beta)}_{\delta)}
-\frac{1}{2}\Phi^{(\alpha}_{\ph{a}(\gamma}\widetilde{Y}_{\delta)}^{\ph{d}\beta)} \label{18}\\
&+\frac{1}{2} g^{\alpha\beta}\Phi_{(\gamma}^{\ph{c}\;\kappa}\widetilde{Y}_{\delta)\kappa}
-\frac{1}{12}\delta^{(\alpha}_{(\gamma}\Phi^{\beta)\kappa}\widetilde{Y}_{\delta)\kappa}
-\frac{1}{4}\Phi_{(\gamma}^{\ph{c}\;\kappa}\delta^{(\alpha}_{\delta)} \widetilde{Y}^{\beta)}_{\ph{a}\;\kappa}=0. \nonumber
\end{align}
Analogously, Eq. (\ref{14}) can be written as 
\begin{multline}
J^{\kappa}\widetilde{C}_{\kappa(\mu\nu)}^{\ph{k}\ph{m}\ph{n}(\alpha}g^{\beta\gamma)}
-J^{(\alpha}\widetilde{C}^{\beta\ph{m}\ph{n}\gamma)}_{\ph{b}(\mu\nu)}
+\frac{1}{2}J^{\kappa}\epsilon_{(\mu\vert\ph{a}\kappa}^{\ph{m}(\alpha\vert\ph{k}\delta}\Phi_{\delta\vert\nu)}g^{\vert\beta\gamma)}\\
+J^{\kappa}\widetilde{C}_{\kappa\ph{a}\ph{b}(\mu}^{\ph{k}(\alpha\beta}\delta^{\gamma)}_{\nu)}
-\frac{1}{2}J^{\kappa}\epsilon^{\delta \ph{f}\ph{k}\;(\alpha}_{\ph{a}(\mu\vert\kappa}\Phi_{\delta}^{\ph{d}\beta}\delta^{\gamma)}_{\vert\nu)}=0.\label{21}
\end{multline}
Now, let us consider the two possible algebraic forms for the KY tensor, null and non-null. These possibilities will be considered separately in what follows.

\subsection{Null Killing-Yano Tensor}\label{SubSec.KYNull}

In what follows we will consider that the KY tensor is a null bivector, so that there exists a null frame such that  
$\bl{Y} = \bl{\ell}\wedge(\bl{m}+\bar{\bl{m}})$, so that its Hodge dual is $\widetilde{\bl{Y}}= i\bl{\ell}\wedge(\bl{m}-\bar{\bl{m}})$. In this case the integrability condition of the KY tensor implies that Weyl tensor is of Petrov type $N$ (or more special, namely $O$), i.e. the only Weyl scalar that can be different from zero is $\Psi_4$, as explained in the previous section. Hence, the Weyl tensor can be written as \cite{Bat-Book}:
\begin{equation}\label{WeylN}
  C_{\mu\nu\alpha\beta} = 4\Psi_4 \,\ell_{[\mu}m_{\nu]} \ell_{[\alpha}m_{\beta]} +  4\bar{\Psi}_4 \,\ell_{[\mu}\bar{m}_{\nu]} \ell_{[\alpha}\bar{m}_{\beta]}\,,
\end{equation}
where $\bar{\Psi}_4$ stands for the complex conjugate of $\Psi_4$. In addition, several components of the trace-less part of the Ricci tensor vanish, in accordance with Eq. (\ref{RicciNull}). The only components that can, in principle, be different from zero are 
$$\Phi_{nn}\,,\;\Phi_{nm}\,,\; \Phi_{n\bar{m}}\,,\; \Phi_{mm}\,,\;  \Phi_{\bar{m}\bar{m}}\,,\; \Phi_{\ell n}\,,\; \Phi_{m\bar{m}}    \,. $$
In addition, the following constraints must hold:
\begin{equation}\label{ConstraintsPhi}
  \left\{
    \begin{array}{ll}
      \Phi_{n\bar{m}} = - \Phi_{nm}\,,\\
      \Phi_{mm} =  \Phi_{\bar{m}\bar{m}} = -2\Phi_{m\bar{m}}=-2\Phi_{\ell n}\,,
    \end{array}
  \right.
\end{equation}
Thus, at the end of the day just three degrees of freedom are left for $\Phi_{\alpha\beta}$, namely $\Phi_{nn}$, $\Phi_{nm}$, and $\Phi_{\ell n}$. 
Similarly, contracting Eq. (\ref{18}) with $n_{\alpha}n_{\beta}m^{\gamma}m^{\delta}$ and $m_{\alpha}m_{\beta}m^{\gamma}n^{\delta}$ leads to $\Phi_{nm}=0$ and $\Phi_{mm}=0$, respectively. Then, taking Eq. (\ref{ConstraintsPhi}) into consideration, it follows that $\Phi_{n\bar{m}}$, $\Phi_{\bar{m}\bar{m}}$, $\Phi_{\ell n}$, and $\Phi_{m\bar{m}}$ are also zero. Hence, the only component of $\Phi_{\alpha\beta}$ that can be different from zero is $\Phi_{nn}$.  Finally, contracting  Eq. (\ref{18}) with $n_{\alpha}n_{\beta}n^{\gamma}m^{\delta}$, we obtain
\begin{equation}
\Psi_{4}+\frac{1}{2}\Phi_{nn}=0.\label{24}
\end{equation}
Therefore, $\Phi_{nn}$ vanishes if, and only if, $\Psi_4$ vanish. Thus, if either $\Phi_{nn}$ or $\Psi_4$ vanish then the spacetime is maximally symmetric, in which case the conserved quantity $Q_{\bl{Y}}$ is useless. In particular, if the spacetime is Einstein, namely if  $\Phi_{\alpha\beta}$ vanish identically then $\Psi_4$ vanishes and we have the trivial case.

Concerning the condition (\ref{21}), contracting it with $m_{\alpha}m_{\beta}m_{\gamma}n^{\mu}\bar{m}^{\nu}$, we obtain $J_\ell\Psi_{4}=0$, where it has been used that $\Psi_4$ is real, which is a consequence of Eq. \eqref{24}. Similarly, contracting with $n_{\alpha}n_{\beta}n_{\gamma}\bar{m}^{\mu}\bar{m}^{\nu}$, $n_{\alpha}n_{\beta}n_{\gamma}\ell^{\mu}m^{\nu}$ and $m_{\alpha}m_{\beta}m_{\gamma}n^{\mu}n^{\nu}$ implies that $J_{n}\Psi_{4}=0$, $(J_{m}+J_{\bar{m}})\Psi_{4}=0$ and $J_{m}\Psi_{4}=0$, respectively. Therefore, the constraint (\ref{21}) leads to
\begin{equation}
J_{\alpha}\Psi_{4}=0\,,\label{26}
\end{equation}
meaning that either $\Psi_4 =0$, which again lead to the trivial case of a maximally symmetric spacetime, or $J_{\alpha}=0$, which means that $\bl{Y}$ is covariantly constant. Indeed, the KY equation can equivalently be written as $\nabla_{\alpha}Y_{\mu\nu} =\nabla_{[\alpha}Y_{\mu\nu]}$. Thus, if $J_\alpha$ vanishes it follows that $\nabla^\alpha\widetilde{Y}_{\alpha\beta}=0$, which is equivalent to the condition $\nabla_{[\alpha}Y_{\mu\nu]}=0$, which implies that $\bl{Y}$ is covariantly constant.

However, if $\bl{Y}$ is covariantly constant so is its Hodge dual $\widetilde{\bl{Y}}$. Particularly, this implies that  $\widetilde{\bl{Y}}$ is also a KY tensor, so that it makes sense to suppose that the scalar $Q_{\widetilde{\bl{Y}}}$ is conserved, although this is not a necessary requirement as it is independent from the requirement that $Q_{\bl{Y}}$ is conserved.  Nevertheless, if besides the conservation of $Q_{\bl{Y}}$ we also assume that  $Q_{\widetilde{\bl{Y}}}$ is conserved, it follows that the condition (\ref{18}) must also hold if we replace $\bl{Y}$ by $\widetilde{\bl{Y}}$. Performing this replacement and then contracting Eq. (\ref{18}) with $n_\alpha n_\beta n^\gamma \bar{m}^\delta $, we end up with the constraint
\begin{equation}
\Psi_{4}-\frac{1}{2}\Phi_{nn}=0.\label{24-2}
\end{equation}
Composing Eqs. (\ref{24}) and (\ref{24-2}) lead us to the conclusion that $\Psi_4$ and $\Phi_{nn}=0$, which then imply that the spacetime is maximally symmetric, in which case the conserved charges are useless.

Summing up, in order for the conserved charge $Q_{\bl{Y}}$ be nontrivial for the case of a KY tensor whose algebraic type is null, the Weyl tensor must be Petrov type $N$ and the only component of $\Phi_{\alpha\beta}$ that can be different from zero is $\Phi_{nn}$. In addition, the KY tensor must be covariantly constant. Due to the latter fact, it follows that $\widetilde{Y}$ is also a KY tensor. If we further impose that $Q_{\widetilde{\bl{Y}}}$ is conserved, in addition to $Q_{\bl{Y}}$,  we conclude that the spacetime is maximally symmetric and the conserved charges are useless.

\subsection{Non-null Killing-Yano Tensor}\label{SubSec.KY.NonNull}

Now, let us assume that the Killing-Yano tensor is non-null, which means that there exists some null frame such that $\bl{Y} = f \,\bl{\ell}\wedge \bl{n} + i h  \,\bl{m} \wedge \bar{\bl{m}}$, where $f$ and $h$ are real functions that cannot vanish simultaneously. The Hodge dual of the KY tensor is then given by
$\widetilde{\bl{Y}} = h \,\bl{\ell}\wedge \bl{n} - i f  \,\bl{m} \wedge \bar{\bl{m}}$. As discussed in Sec. \ref{Sec.KY}, in this case the integrability condition of the KY tensor implies that the only Weyl scalar that can be different from zero is $\Psi_2$, so that the Weyl tensor can be written as follows \cite{Bat-Book}:
\begin{align} 
  C_{\mu\nu\alpha\beta} &= 
(\Psi_2+\bar{\Psi}_2) \,\left( \ell_{[\mu}n_{\nu]} \ell_{[\alpha}n_{\beta]} +  m_{[\mu}\bar{m}_{\nu]} m_{[\alpha}\bar{m}_{\beta]} \right) \nonumber\\
&-(\Psi_2-\bar{\Psi}_2) \, \left( \ell_{[\mu}n_{\nu]} m_{[\alpha}\bar{m}_{\beta]}  +  m_{[\mu}\bar{m}_{\nu]}  \ell_{[\alpha}n_{\beta]} \right) \nonumber \\
&-\Psi_2 \left( \ell_{[\mu}m_{\nu]} n_{[\alpha}\bar{m}_{\beta]} +  n_{[\mu}\bar{m}_{\nu]} \ell_{[\alpha}m_{\beta]} \right) \nonumber\\
&- \bar{\Psi}_2 \left( \ell_{[\mu}\bar{m}_{\nu]} n_{[\alpha}m_{\beta]} +  n_{[\mu}m_{\nu]} \ell_{[\alpha}\bar{m}_{\beta]} \right)\,. \label{WeylD}
\end{align}
In addition, the following components of the trace-free part of the Ricci tensor must vanish due to the fact that $\bl{Y}$ is a KY tensor:
\begin{equation}\label{RicciD}
  \Phi_{\ell m}=\Phi_{\ell \bar{m}}= \Phi_{nm}= \Phi_{n \bar{m}} =0\,.
\end{equation}

Now, taking Eqs. (\ref{WeylD}) and (\ref{RicciD}) into consideration, we are ready to analyse Eq. (\ref{18}), which is necessary for $Q_{\bl{Y}}$ be conserved. Contracting (\ref{18}) with $n_{\alpha}n_{\beta}n^{\gamma}\ell^{\delta}$,  $\ell_{\alpha}\ell_{\beta}\ell^{\gamma}n^{\delta}$, $m_{\alpha}m_{\beta}m^{\gamma}\bar{m}^{\delta}$, and $\bar{m}_{\alpha}\bar{m}_{\beta}\bar{m}^{\gamma}m^{\delta}$ we obtain respectively:
\begin{equation}\label{RicciD2}
  \Phi_{nn}=0\,,\;\Phi_{\ell\ell}=0\,,\;\Phi_{mm}\,,\;\Phi_{\bar{m}\bar{m}}=0 \,.
\end{equation}
Since the trace-free condition obeyed by $\bl{\Phi}$ means that $\Phi_{\ell n}=\Phi_{m\bar{m}}$, it follows that both components  $\Phi_{\ell n}$ and  $\Phi_{m\bar{m}}$ represent the same degree of freedom. Hence, from  Eqs. (\ref{RicciD}) and (\ref{RicciD2}) one concludes that only one degree of freedom of $\bl{\Phi}$ can be different from zero, namely $\Phi_{\ell n}$.

Then, contracting Eq. (\ref{18}) with $n_{\alpha}n_{\beta}\ell^{\gamma}\ell^{\delta}$ and $m_{\alpha}m_{\beta}\bar{m}^{\gamma}\bar{m}^{\delta}$, we arrive at the following relations respectively
\begin{eqnarray}
h\, Re\{\Psi_{2}\}+f\, Im\{\Psi_{2}\}+\frac{1}{3}h\, \Phi_{n\ell}=0\;,\nonumber\\
-f\, Re\{\Psi_{2}\}+h\, Im\{\Psi_{2}\}+\frac{1}{3}f\,\Phi_{m\bar{m}}=0\;.
\end{eqnarray}
Finally, using $\Phi_{n\ell}=\Phi_{m\bar{m}}$, we conclude that
\begin{equation}\label{WeylRicciD}
\Psi_{2} =\frac{1}{3}\,\frac{f-ih}{f+i h}\,\Phi_{n\ell}\,. 
\end{equation}
Thus, if the spacetime is Einstein, i.e. if $\Phi_{\alpha\beta}=0$, then $\Psi_{2}$ vanishes. The latter, in turn, is the unique Weyl scalar that can be different from zero, so that we conclude that the whole Weyl tensor vanishes. Hence, if the spacetime is Einstein it will also be conformally flat and these two conditions means that the spacetime is maximally symmetric, so that the conserved quantity $Q_{\bl{Y}}$ is trivial.

Regarding the constraint (\ref{21}), one can check that it boils down to 
\begin{equation*}
J_{\alpha}\,\Psi_{2}=0 \,,
\end{equation*}
where Eq. (\ref{WeylRicciD}) has been used. 
Hence, either the space is maximally symmetric (if $\Psi_2=0$, which then implies $\Phi_{\alpha\beta}=0$), or the KY tensor is covariantly constant (if $J_{\alpha}=0$). Thus, the only non-trivial case in which $Q_{\bl{Y}}$ is conserved for a non-null KY tensor is when this tensor is covariantly constant, 
the Weyl tensor is of Petrov type $D$ and with the only nonvanishing components of $\Phi_{\alpha\beta}$ being $\Phi_{\ell n} = \Phi_{m\bar{m}}$. Furthermore, the relation between $\Psi_2$ and $\Phi_{\ell n}$ given in Eq. (\ref{WeylRicciD}) must hold. These are quite restrictive conditions.

Now, since $\bl{Y}$ is covariantly constant, it follows that its Hodge dual is also a KY tensor. Then we can require that $Q_{\widetilde{\bl{Y}}}$ is also conserved along the solutions of the MPD equation, although it is worth pointing out that this is an independent requirement. Comparing the expressions for $\bl{Y}$ and $\widetilde{\bl{Y}}$,
\begin{equation*}
  \left\{
     \begin{array}{ll}
       \bl{Y} =  f \,\bl{\ell}\wedge \bl{n} + i h  \,\bl{m} \wedge \bar{\bl{m}} \\
       \widetilde{\bl{Y}} =  h \,\bl{\ell}\wedge \bl{n} - i f  \,\bl{m} \wedge \bar{\bl{m}}
     \end{array}
   \right.\,,
\end{equation*}
we note that one $\widetilde{\bl{Y}}$ can be obtained from $\bl{Y}$ by making the changes $f\rightarrow h$ and $h\rightarrow -f$. Thus, since Eq. (\ref{WeylRicciD}) must hold in order to guarantee that $Q_{\bl{Y}}$ is conserved, it follows that the analogous condition
\begin{equation}\label{WeylRicciD2}
\Psi_{2} =\frac{1}{3}\,\frac{h+if}{h-i f}\,\Phi_{n\ell} 
\end{equation}
must hold in order to assure the conservation of $Q_{\widetilde{\bl{Y}}}$. Hence, assuming that the scalars $Q_{\bl{Y}}$ and $Q_{\widetilde{\bl{Y}}}$ are both conserved, it follows that Eqs. (\ref{WeylRicciD}) and (\ref{WeylRicciD2}) hold simultaneously. Equating both expressions for $\Psi_2$ and assuming that $\Phi_{\ell n}\neq 0$, so that the spacetime is nontrivial, lead us to the condition
$$  \frac{h+if}{h-i f} = \frac{f-ih}{f+i h} \; \Rightarrow\; f^2 + h^2 = 0 \,.$$ 
Since $f$ and $h$ are real functions, the unique solution for the latter constraint turns out to be the trivial one, $f=h=0$, which is unacceptable, since by hypothesis $\bl{Y}$ is a nonvanishing KY tensor. Thus, we conclude that the only case in which $Q_{\bl{Y}}$ and $Q_{\widetilde{\bl{Y}}}$ are both conserved is when $\Phi_{\ell n}=0$, which then implies $\Psi_2=0$. This means that the spacetime is maximally symmetric and, therefore, the conserved scalars of interest are useless.

\subsection{Physical Restrictions by Energy Conditions}

As we have just seen, the integrability conditions for the KY tensor along with the additional conditions required for $Q_{\bl{Y}}$ be conserved inflict huge restrictions over the Weyl and Ricci tensors. In the present subsection we shall make use of Einstein's equation to convert the restrictions over the Ricci tensor onto constraints over the energy-momentum tensor of the matter on the background. More precisely, we shall analyse whether the weak energy condition (WEC) holds or not. Here we will assume that the spacetime is not maximally symmetric, which means that we are requiring that just $Q_{\bl{Y}}$ is conserved, while $Q_{\widetilde{\bl{Y}}}$ is not a conserved scalar, otherwise $\Phi_{\mu\nu}$ would vanish identically and the calculations below would be senseless.

In suitable units, Einstein's equation reads $G_{\mu\nu}=T_{\mu\nu}$, where $T_{\mu\nu}$ is the energy-momentum tensor of the background matter. This can be equivalently written as 
$$ T_{\mu\nu} = \Phi_{\mu\nu} - \frac{R}{4}g_{\mu\nu}\,. $$
The weak energy condition then amounts to the constraint $T_{\mu\nu} Z^\mu Z^\nu \geq 0$ for any time-like vector field $Z^\mu$, which means that the energy density of the matter is not negative as measured by an arbitrary observer. Writing the vector field $\bl{Z}$ in terms of the null tetrad frame we have
$$ \bl{Z} = Z_n\,\bl{\ell} + Z_\ell\,\bl{n} -\, Z_{\bar{m}}\,\bl{m} -\,Z_{m}\,\bar{\bl{m}}.  $$
The WEC then reads
\begin{equation}\label{WECZ}
  \Phi_{\mu\nu}Z^\mu Z^\nu  - \frac{R}{2} \left( Z_n Z_\ell -Z_{m} Z_{\bar{m}}  \right) \geq 0  \,,
\end{equation}
for any vector $\bl{Z}$ such that $ Z_n Z_\ell > Z_{m} Z_{\bar{m}} $. 
Since most of the components of $\Phi_{\mu\nu}$ vanish when $Q_{\bl{Y}}$ is conserved, the above restriction becomes simpler to be analysed. In what follows let us consider the two possible algebraic types of the KY tensor separately.

When the KY tensor is type null, the only component of $\Phi_{\mu\nu}$ that can be different from zero is $\Phi_{nn}$, so that Eq. (\ref{WECZ}) becomes
$$  \Phi_{nn}Z_\ell Z_\ell  - \frac{R}{2} \left( Z_n Z_\ell -Z_{\bar{m}} Z_{m}  \right) \geq 0 \,. $$
Defining $\zeta \equiv \left( Z_n Z_\ell -Z_{\bar{m}} Z_{m}  \right)/(Z_\ell^2)$, it follows that the time-like condition reads $\zeta >0 $, so that the WEC becomes 
$$   \Phi_{nn} \geq  \frac{R}{2} \zeta \;,\;\; \text{ for all }\zeta>0\,. $$
This is possible only if $\Phi_{nn}\geq 0$ and $R\leq0$. Thus, besides the geometrical restrictions found in subsection \ref{SubSec.KYNull}, there exists the physical restriction that the Ricci scalar cannot be positive whereas the component $\Phi_{nn}$ cannot be negative. Otherwise the background spacetime is not generated by a physically reasonable matter.

Now, let us consider that the KY tensor has a non-null algebraic type, in which case the only components of $\Phi_{\mu\nu}$ that can be different from zero are $\Phi_{\ell n} = \Phi_{m\bar{m}}$, so that Eq. (\ref{WECZ}) becomes
$$  \Phi_{\ell n} (Z_\ell Z_n + Z_{\bar{m}} Z_{m} ) - \frac{R}{4} \left( Z_n Z_\ell -Z_{\bar{m}} Z_{m}  \right) \geq 0  \,. $$
Since the time-like condition for $\bl{Z}$ reads 
$$ Z_n Z_\ell > Z_{m} Z_{\bar{m}} = |Z_m|^2 \,,$$ 
it follows that $Z_n Z_\ell$ is positive and, therefore, defining $\xi \equiv (Z_n Z_\ell - |Z_m|^2)/(Z_n Z_\ell +|Z_m|^2 )$, it follows that $\xi$ is positive, so that the WEC for spacetimes with conserved $Q_{\bl{Y}}$ for a non-null KY tensor is given by
$$  \Phi_{\ell n}  \geq  \frac{R}{4} \,\xi\;,\;\; \text{ for all }\xi>0\,   \,. $$
This, in turn, implies that $\Phi_{\ell n}$ cannot be negative and the Ricci scalar cannot be positive.

\section{Looking for Explicit Examples}\label{Sec.Examples}

The aim of the present section is to find non-trivial examples of spacetimes obeying the several restrictions necessary in order to assure the conservation of $Q_{\bl{Y}}$. We shall start analysing the case in which the KY tensor is null and then consider the non-null case.

\subsection{An Example with a Null KY Tensor}\label{SubSec.ExampleN}

As argued in Sec. \ref{Sec.KY},  when the algebraic type of the KY tensor is null the Weyl tensor must be type $N$ according to the Petrov classification. A well-known class of type $N$ spacetimes is given by the so-called $pp-$wave metrics. These spacetimes are generally associated to gravitational radiation and are geometrically defined as the ones possessing a covariantly constant null vector field. Their line elements are given by
\begin{equation}
ds^{2}= 2 F(u,z,\bar{z}) \,du^{2} + 2 \,du dr - 2\, dz\,d\bar{z}\,,
\end{equation}
where $u$ and $r$ are real coordinates, whereas $z$ is a complex coordinate with $\bar{z}$ being its complex conjugate. $F$ is an arbitrary real function of the coordinates $u$, $z$, and $\bar{z}$.  A null tetrad frame is then given by 
$$ \bl{\ell} = \pd_r\,,\; \bl{n} = \pd_{u} - F\pd_r\,,\; \bl{m}=\pd_z\,,\; \bar{\bl{m}}=\pd_{\bar{z}}\,.  $$
The null vector field $\bl{\ell}$ is the covariantly constant vector that characterizes a $pp-$wave spacetime. In this frame the unique Weyl scalar that is different from zero is 
\begin{equation}\label{Psi4pp}
  \Psi_4 = - \pd_{\bar{z}}\pd_{\bar{z}} F\,,
\end{equation}
whereas the only component of the Ricci tensor that is different from zero, in this null frame, is 
$$ R_{nn} = \Phi_{nn} =  2\pd_{z}\pd_{\bar{z}} F \,. $$
The null bivector $\bl{Y} = \bl{\ell}\wedge (\bl{m} + \bar{\bl{m}})$ is covariantly constant and, therefore, is also a KY tensor. Thus, out of the restrictions necessary in order to $Q_{\bl{Y}}$ be conserved, the only that remains to be met is the one given in Eq. (\ref{24}), namely $\Psi_{4}+\frac{1}{2}\Phi_{nn}=0$. Imposing the latter equation to hold, lead us to the partial differential equation $\pd_{\bar{z}}\pd_{\bar{z}} F=\pd_{z}\pd_{\bar{z}} F$, whose general solution is
\begin{equation}\label{F}
  F(u,z,\bar{z}) =  F_1(u,z+\bar{z}) + F_2(u,z) \,,
\end{equation}
where $F_1$ and $F_2$ are general real functions of their arguments. Note however, that taking the complex conjugate of the equation (\ref{24}) it follows that $\Psi_4$ must be a real function, since the Ricci tensor is clearly real and the null vector $\bl{n}$ is also real. Therefore, from Eq. (\ref{Psi4pp}), it follows that 
\begin{equation*}\label{Psi4pp2}
  \Psi_4 = \bar{\Psi}_4 \; \Rightarrow\;  \pd_{\bar{z}}\pd_{\bar{z}} F =  \pd_{z}\pd_{z}F \,,
\end{equation*}
This condition, along with Eq. (\ref{F}) implies that the function $F$ must have the form
\begin{equation}\label{F2}
  F(u,z,\bar{z}) =  F_3(u,z+\bar{z}) \,,
\end{equation} 
where $F_3$ is an arbitrary real function of $u$ and $z+\bar{z}$. This choice of function $F$ leads to the most general $pp-$wave spacetime that such that the scalar 
\begin{equation*}
   Q_{\bl{Y}} = S^{\mu\nu} \widetilde{Y}_{\mu\nu} = 2i\left(  S_{\ell m} - S_{\ell\bar{m}} \right) =
2i\left(  S_{rz} - S_{r\bar{z}} \right)
\end{equation*}
\\
is conserved along the solutions of the MPD equations, where in the last equality it has been used that $\bl{Y}$ is the bivector $\pd_r\wedge (\pd_z + \pd_{\bar{z}})$.

However, it turns out that the bivector $\widetilde{\bl{Y}} = i\pd_r\wedge (\pd_z - \pd_{\bar{z}}) $ is also a KY tensor (actually it is covariantly constant). Imposing the scalar $Q_{\widetilde{\bl{Y}}}$ to be conserved we would find from R\"{u}diger's conditions that the function $F$ appearing in the line element should have the form
\begin{equation}\label{F3}
  F(u,z,\bar{z}) =  F_4(u,z-\bar{z}) \,. 
\end{equation} 
Note that Eqs. (\ref{F2}) and (\ref{F3}) hold simultaneously only if $F$ is a function of $u$ alone, $F = F(u)$, in which case the spacetime would be maximally symmetric, in accordance with what has been obtained in Sec. \ref{SubSec.KYNull} when the constancy of $Q_{\bl{Y}}$ and $Q_{\widetilde{\bl{Y}}}$
are imposed simultaneously.

\subsection{Seeking for an Example with a Non-Null KY Tensor}

Since the most general metric of Petrov type $D$ possessing a covariantly constant bivector $\bl{Y}$ is not available in the literature and certainly is quite hard to find it, here we will start with the most general type $D$ spacetime possessing a KY tensor and two commuting Killing vectors. The latter class of spacetimes is physically relevant due to the fact that a star that have attained the equilibrium should be stationary and axissymmetric, which geometrically means that there exists Killing vector fields $\pd_\tau$ and $\pd_\varphi$. Moreover, the existence of a KY tensor along with the two Killing vectors assure the integrability of the geodesic motion. In particular, Kerr metric is a member of this class of spacetimes. The most general metric possessing these features has been obtained in Ref. \cite{AnabalonBatista} and is given by:
\begin{align*}
  ds^2 &= S\,\Big[   \frac{A_2 \Delta_2 }{(x^2+y^2)^2}(dt + x^2d\varphi^2)^2 - \frac{dy^2}{\Delta_2}\\
&-  \frac{A_1 \Delta_1 }{(x^2+y^2)^2}(dt - y^2d\varphi^2)^2 - \frac{dx^2}{\Delta_1}   \Big]\,,
\end{align*}
where $\Delta_1$ and $\Delta_2$ are arbitrary functions whereas $A_1$, $A_2$, and $S$ are the functions given by
\begin{align*}
  A_1 = &\frac{x^2}{(b_1 x^2 + \eta_1)(b_2 x^2 + \eta_2)} \,,
\\  A_2 = & \frac{y^2}{( \eta_1- b_1 y^2 )(b_2 y^2 - \eta_2)} \\
  S = & \frac{b_3 x^2 + \eta_3}{b_1 x^2 + \eta_1} + \frac{b_3 y^2 - \eta_3}{ \eta_1 - b_1 y^2 } \,,
\end{align*}
where the $b$'s and $\eta$'s are arbitrary constants. The null tetrad frame aligned with the principal null directions of the Weyl tensor is given by
\begin{align*}
  \bl{\ell} &= \frac{1}{\sqrt{2 S \Delta_2 }}\,\left( \frac{y^2}{\sqrt{A_2}} \pd_t + \frac{1}{\sqrt{A_2}}\pd_\varphi -\Delta_2\pd_y  \right)\,, \\
  \bl{n} &=  \frac{1}{\sqrt{2 S \Delta_2 }}\,\left( \frac{y^2}{\sqrt{A_2}} \pd_t + \frac{1}{\sqrt{A_2}}\pd_\varphi +\Delta_2\pd_y  \right) \,, \\
\bl{m} &=    \frac{1}{\sqrt{2 S \Delta_1 }}\,\left( \frac{x^2}{\sqrt{A_1}} \pd_t - \frac{1}{\sqrt{A_1}}\pd_\varphi +i\,\Delta_1\pd_x  \right) \,, \\
\bl{\bar{m}} &=  \frac{1}{\sqrt{2 S \Delta_1 }}\,\left( \frac{x^2}{\sqrt{A_1}} \pd_t - \frac{1}{\sqrt{A_1}}\pd_\varphi -i\,\Delta_1\pd_x  \right)\,.    \\
\end{align*}
The KY tensor is given by
$$ \bl{Y} = f\bl{\ell}\wedge \bl{n} + i h \bl{m}\wedge \bar{\bl{m}} \,,$$ 
where 
$$ f = -\sqrt{\frac{b_2 x^2 + \eta_2}{b_1 x^2 + \eta_1 }}  \; \text{and}\; h = \sqrt{\frac{b_2 y^2 - \eta_2}{ \eta_1 - b_1 y^2 }} $$
Using this frame it follows that the only Weyl scalar that is different from zero is $\Psi_2$, whereas the components of $\Phi_{\mu\nu}$ in this frame are all vanishing apart from $\Phi_{\ell n}$ and $\Phi_{m\bar{m}}$, where it is worth recalling that the trace-free condition implies 
$\Phi_{\ell n}=\Phi_{m\bar{m}}$. Then, the only constraints that remain to be imposed in order to assure that $Q_{\bl{Y}}$ is conserved along the solutions of the MPD equations are Eq. (\ref{WeylRicciD}), which connects $\Psi_2$ and $\Phi_{\ell n}$, and the requirement that $\bl{Y}$ must be covariantly constant. In particular, imposing the latter constraint we find that either $b_3/\eta_3 = b_1/\eta_1$ or $b_2/\eta_2 = b_1/\eta_1$, but the former option leads to a vanishing $S$ and, therefore, a vanishing metric, which is senseless. Thus, let us consider $b_1/\eta_1 = b_2/\eta_2 $. However, in this case either $A_1$ or $A_2$ become negative, so that the signature ceases to the Lorentzian, i.e. the space is nonphysical. Thus, for the broad class of spacetimes considered here there exist no example in which the scalar $Q_{\bl{Y}}$ is conserved along the solutions of the MPD equations.

\section{Conclusions}\label{Sec.Conclusions}

We have proved that the integrability condition of the KY tensor along with the constraints necessary for $Q_{\bl{Y}}$ to be conserved imply, actually, that the bivector $\bl{Y}$ should be more than a KY tensor, it must be a covariantly constant tensor. In addition, we have proved that if the background is an Einstein space then the conservation of the scalar $Q_{\bl{Y}}$ implies that the spacetime must be maximally symmetric, i.e. trivial. This is a great improvement on the understanding of the conserved quantity introduced by R\"{u}diger in Ref. \cite{Rudiger1}. There exists several physically relevant spacetimes possessing KY tensors, as exemplified by Kerr-NUT-(A)dS and Kerr-Newman metrics. However, backgrounds possessing covariantly constant bivectors and with physical interest are much more rare. This greatly undermines the usefulness of the conserved scalar $Q_{\bl{Y}}$. Moreover, once $\bl{Y}$ is constant, it follows that its Hodge dual is also constant and, therefore, is also a KY tensor. Hence it is natural to demand that the scalar constructed from $\widetilde{\bl{Y}}$, namely $Q_{\widetilde{\bl{Y}}}$, should also be constant. In this case, it turns out that the spacetime must be maximally symmetric, which, in turn, means that these conserved scalars are useless for the integration of MPD equations, since in these spaces full integrability can already be attained by means of the Killing vector fields. However, it is worth pointing out that in spite of being reasonable to require that $Q_{\bl{Y}}$ and $Q_{\widetilde{\bl{Y}}}$ are both conserved, this is not necessary. Rather, we could be interest on finding spaces in which just $Q_{\bl{Y}}$ is conserved. In the present article we have proved that there exist spacetimes obeying the latter condition, but they form a very narrow class of metrics. Indeed, we have proved that besides having the covariantly constant bivector $\bl{Y}$, these spacetimes must have Weyl tensors are either of Petrov type $N$, when $\bl{Y}$ is a null bivector, or type $D$, when $\bl{Y}$ is non-null. Moreover, using the null tetrad frame adapted to the covariantly constant bivector, we have seen that the trace-less part of the Ricci tensor must have just one non-vanishing degree of freedom and this degree of freedom is connected to the only Weyl scalar that can be different from zero, see Eqs. (\ref{24}) and (\ref{WeylRicciD}). In particular, we have provided one explicit example in Sec. \ref{SubSec.ExampleN}.

The scenario of greater physical interest for the use of the MPD equations is given by a test particle moving in empty space around some celestial body, so that the energy-momentum tensor of the background matter vanishes in the region of interest. Einstein's equation then implies that the trace-less part of the Ricci tensor vanishes (even allowing the existence of a cosmological constant), in which case the conservation of $Q_{\bl{Y}}$ implies that the spacetime is maximally symmetric. Therefore, the conserved quantity $Q_{\bl{Y}}$ is useless in most scenarios of physical relevance.

As a final comment, it is worth pointing out that the idea of R\"{u}diger's article was to look for a scalar linear on the momenta that is conserved for an arbitrary theory yielding MPD equations and adopting Tulczyjew supplementary condition $S^{\alpha\beta}P_{\beta}=0$. In this broad scenario, R\"{u}diger obtained that $Q_{\bl{Y}} = S^{\mu\nu} \widetilde{Y}_{\mu\nu}$ is conserved provided that $\bl{Y}$ is a KY tensor and conditions (\ref{11}) and (\ref{12}) are obeyed. However, in some specific theories these extra conditions might not be necessary and even other conserved scalars might exist. As an example, let us consider the Lagrangian formulation of the spinning particle theory \cite{Hanson:1974qy,armaza}, with the following specific Lagrangian:
$$ L = a\,V^\mu V_{\mu} + b\, \sigma^{\mu\nu}\sigma_{\mu\nu} \,,$$ 
where $a$ and $b$ are nonvanishing constants $V^\mu = dx^\mu/d\tau$. The momenta are then defined by \cite{armaza}:
$$ P^\mu = -\frac{\pd L}{\pd V_\mu}  \;\; \text{and} \;\; S^{\mu\nu} =  -\frac{\pd L}{\pd \sigma_{\mu\nu}} \,, $$
which for this particular Lagrangian yields
$$ P^\mu = -2 a V^\mu  \;\; \text{and} \;\; S^{\mu\nu} = -2b \sigma^{\mu\nu} \,. $$
The field equations for this Lagrangian are MPD equations. Since in this case $\bl{P}\propto \bl{V}$, it follows from Eq. (\ref{MPDeq}) that $\dot{S}^{\mu\nu}=0$, so that the conservation of $Q_{\bl{B}} = S^{\mu\nu} \widetilde{B}_{\mu\nu}$ for some bivector $B^{\mu\nu}$ reads
\begin{equation}\label{QLagrangian}
   \dot{Q}_{\bl{B}} = S^{\mu\nu}V^{\alpha}\nabla_\alpha \widetilde{B}_{\mu\nu} =0\,.
\end{equation}
Now, the supplementary condition $S^{\mu\nu}P_\nu = 0$ can be equivalently written as $S^{\mu\nu} = \epsilon^{\mu\nu\gamma\delta} \Sigma_\gamma P_\delta$, where $\Sigma^\alpha P_{\alpha}=0$. Thus, Eq. (\ref{QLagrangian}) reads
\begin{equation}\label{QLagrangian}
   \dot{Q}_{\bl{B}} = -\,\frac{1}{a}  \,\Sigma_\gamma P_\delta P_\alpha \nabla^\alpha B^{\gamma\delta} =0\,.
\end{equation}
Imposing that the above equation holds for an arbitrary $P^\mu$ and an arbitrary $\Sigma^\mu$ orthogonal to $\bl{P}$ we eventually find that 
$\nabla^{(\alpha} B^{\delta)\gamma}$ must vanish, i.e. $\bl{B}$ must be a KY tensor, which agrees with R\"{u}diger's result. But in this specific theory note that no additional condition is required for the conservation of $Q_{\bl{B}}$.

\begin{acknowledgments}
C. B. would like to thank Conselho Nacional de Desenvolvimento Cient\'{\i}fico e Tecnol\'ogico (CNPq) for the partial financial support through the research productivity fellowship. Likewise,  C. B. thanks Universidade Federal de Pernambuco for the funding through Qualis A project and CAPES for the support of the graduation program.  E. B. S. thanks CAPES for the doctor scholarship. We both thank Bruno de S. L. Torres for the valuable interactions at the beginning of this project.  
\end{acknowledgments}

\end{document}